\documentclass[12pt]{iopart}					
\usepackage{graphicx}						

\begin{document}
\title{Controlled double-slit electron diffraction}

\author{Roger~Bach$^1$, Damian~Pope$^2$, Sy-Hwang~Liou$^1$ and Herman~Batelaan$^1$}

\address{$^1$ Department of Physics and Astronomy, University of Nebraska-Lincoln, Theodore P. Jorgensen Hall, Lincoln, NE 68588, USA.}
\address{$^2$ Perimeter Institute for Theoretical Physics, 31 Caroline ST N Waterloo, Ontario, Canada N2L2Y5.}
\ead{hbatelaan@unl.edu}

\begin{abstract}
Double-slit diffraction is a corner stone of quantum mechanics.
It illustrates key features of quantum mechanics: interference and the particle-wave duality of matter.
In 1965, Richard Feynman presented a thought experiment to show these features. Here we demonstrate the full realization
of his famous thought experiment. By placing a movable mask in front of a double-slit to control the transmission
through the individual slits, probability distributions for single- and double-slit arrangements were observed.
Also, by recording single electron detection events diffracting through a double-slit, a diffraction pattern was
built up from individual events.
\end{abstract}
\pacs{03.65.-w, 03.75.-b, 41.75.Fr, 41.85.-p, 42.25.Fx}
\maketitle

\section{Introduction}
Richard Feynman described electron diffraction as a phenomenon ``which has in it the heart of quantum mechanics.
In reality, it contains the \emph{only} mystery''\cite{Feynman65}. He went on to describe a thought experiment for 
which he stated ``that you should not try to set up'' because ``the apparatus would have to be made on an impossibly small
scale to show the effects we are interested in.''  
He used these effects to help illustrate the phenomena of wave-particle duality, which is a postulate that all particles exhibit both wave and particle properties.
The effects he described were: the relations between electron probability distributions from single- and double-slits,
and observation of single particle diffraction.
In this paper we report both control over the individual slits to observe probability distributions from both single- and double-slits, and
the build-up of a diffraction pattern at single electron detection rates to achieve the full realization of Feynman's thought experiment.

The general perception is that the electron double-slit experiment has already been performed. This is true in the sense 
that J\"onsson demonstrated diffraction from single, double, and multiple (up to five) micro-slits\cite{Jonsson61}, but he could not
observe single particle diffraction, nor close individual slits. 
In two separate landmark experiments, individual electron detection was used to build up interference patterns;
however, biprisms were used instead of double-slits\cite{Pozzi76,Tonomura89}.
First, Pozzi recorded the interference patterns at varying electron beam densities.
Then, Tonomura recorded the positions of individual electron detection events as a function of time and used them to build up an interference pattern. 
It is interesting to point out that the build up of a double-slit diffraction pattern; i.e., measured one electron at a time, has been called ``The most beautiful experiment in physics''\cite{Beautiful02,DoubleSlit02},
while the build-up experiment for a true double-slit has, up to now, never been reported.

More recently, electron diffraction was demonstrated with single- and double-slits using Focused Ion Beam (FIB) milled nano-slits\cite{Barwick06,Pozzi07}. 
In addition, one single slit in a double-slit was closed by FIB induced deposition\cite{Pozzi08}. 
This process is not reversible, so observation of the electron probability distribution through both single-slits could 
not be done, and single electron detection was not reported.

Feynman's thought experiment is summarized in figure~\ref{fig1}. The figure is an adaptation from \emph{Feynman Lectures on Physics}, Volume III, figure 1-3,
with the mask, experimental data, and micrographs added.
The thought experiment contained two parts.
The first involved observing probability distributions in three scenarios:
electrons traveling through slit 1 with slit 2 closed ($\textrm{P}_{1}$);
electrons traveling through slit 2 with slit 1 closed ($\textrm{P}_{2}$);
and electrons traveling through both slits ($\textrm{P}_{12}$).
These scenarios illustrate the quantum mechanical superposition principle, i.e., the wave properties, and can be demonstrated with control of the slits (figure~\ref{fig2}).
The second part of the thought experiment was the observation of individual electrons associated with detection ``clicks''. 
This illustrates that a quantum mechanical electron wave can not be thought of 
as comprising multiple electrons, i.e., the particle properties, which can be demonstrated with the build-up of the diffraction pattern (figure~\ref{fig3}).

\section{Experimental Setup}
The experimental setup is shown diagrammatically in figure~\ref{fig1}a.
An electron beam with energy of 600~eV was generated with a thermionic tungsten filament and several electrostatic lenses.
The beam was collimated with a slit of 2 $\mu$m width and 10 $\mu$m height placed at 16.5 cm.
The double-slit was located 30.5 cm from the collimation slit.
The resulting patterns were magnified by an electrostatic quadrupole lens and imaged on a two-dimensional microchannel plate and
phosphorus screen, then recorded with a charge-coupled device camera.

Two methods were used to analyze the images.
To investigate the probability distributions, the images were summed up by adding each frame's intensity, then normalized.
This resulted in a false colour probability distribution (figure~\ref{fig1}~and~\ref{fig2}).
To study the build-up of the diffraction pattern, each electron was localized using a ``blob'' detection scheme\cite{Lindeberg94,Lindeberg98}. Each detection was replaced by a blob,
whose size represents the error in the localization of the detection scheme. The blobs were compiled together to form the electron diffraction patterns (figure~\ref{fig3}).

The collimation slit, double-slit, and mask were made by FIB milling into three 100-nm-thin silicon-nitride membrane windows.
The FIB milling was performed on a 30-keV $\textrm{Ga}^{+}$ system (FEI Strata 200xp).
After milling, each membrane was coated with approximately 2 nm of gold.
The double-slit consists of two 50-nm-wide slits with a center-to-center separation of 280 nm (see inset 1 in figure~\ref{fig1}).
Each slit is 4 $\mu$m tall and has a 150 nm support midway along it's height.
The mask is 5 $\mu$m wide $\! \times \!$ 10 $\mu$m tall (see inset 2 in figure~\ref{fig1}),
and was placed 230 $\mu$m away from the double-slit.
The mask was held securely in a frame that could slide back and forth and was controlled by a piezoelectric actuator.

\section{Results}
\begin{figure}[b]
\centering
\includegraphics[width=\textwidth]{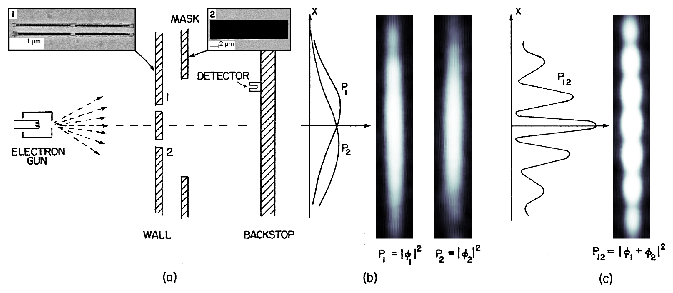}
\caption[Setup]{Simplified setup. \textbf{a}, An electron beam passes through a wall with two slits in it.
A movable mask is positioned to block the electrons, only allowing the ones traversing through slit 1 ($\textrm{P}_{1}$),
slit 2 ($\textrm{P}_{2}$), or both ($\textrm{P}_{12}$) to reach the backstop and detector. \textbf{b,c}, Probability 
distributions are shown, (Experimental in false-colour intensity) for electrons that pass through a single slit (b),
or the double-slit (c). \textbf{Inset 1,2}, Electron micrographs of the double-slit and mask are shown. 
The individual slits are 50 nm wide $\times$ 4 $\mu$m tall with a 150 nm support structure midway along it's height,
and separated by 280 nm. The mask is 5 $\mu$m wide $\times$ 20 $\mu$m tall.
Reprinted from \emph{The Feynman Lectures on Physics}, Volume III, by Richard P. Feyman, Robert B. Leighton, and 
Matthew Sands. Available from Basic Books, an imprint of The Perseus Books Group. Copyright $\copyright$ 2011}
\label{fig1}
\end{figure}

\begin{figure}[tb!]
\centering
\includegraphics[keepaspectratio=true,height=.800\textheight,width=\columnwidth]{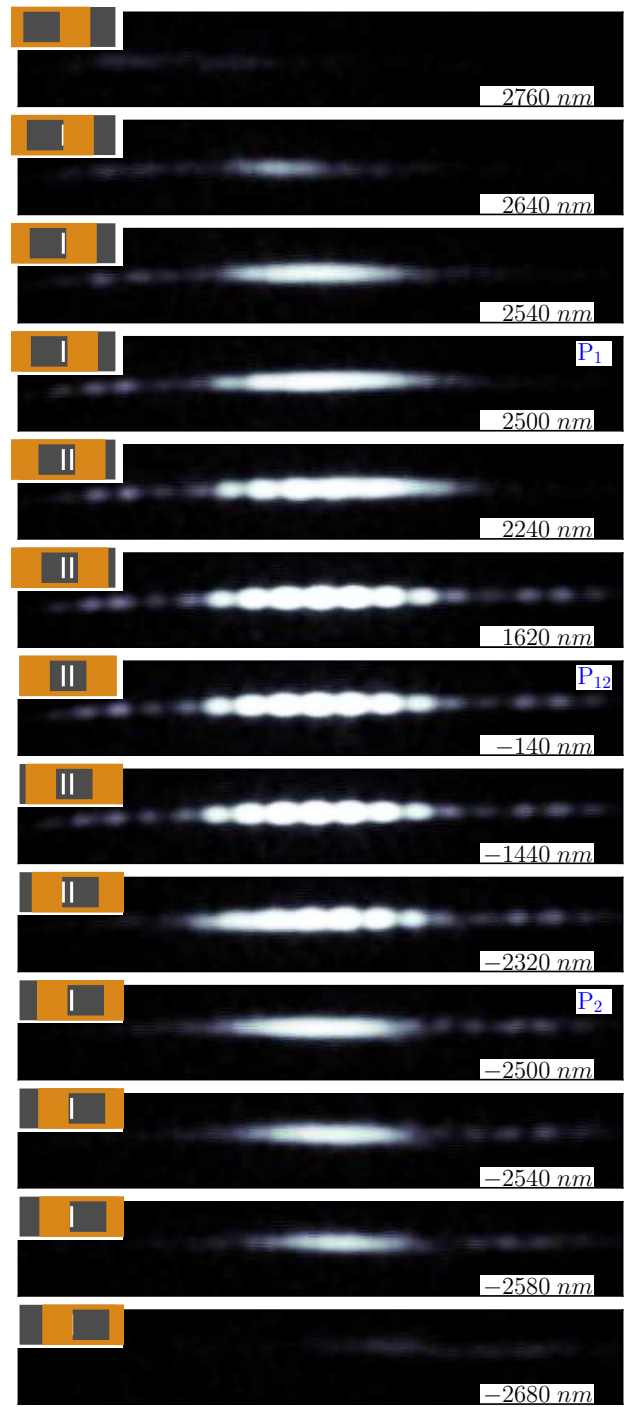}
\caption[Mask movement]{Mask movement. A mask is moved over a double-slit (inset) and the resulting 
probability distributions are shown.  The mask allows the blocking of one slit, both slits, or neither slit in a non
destructive way. The individual slits are 50 nm wide and separated by 280 nm. The mask has a 5 $\mu$m wide
opening. The labeled dimensions are the positions of the center of the mask. $\textrm{P}_{1}$, 
$\textrm{P}_{2}$, and $\textrm{P}_{12}$ are the probability distributions shown in 
figure~\ref{fig1}. (See Supplementary Movie 1 for more positions of the mask.)}
\label{fig2}
\end{figure}

The movable mask was placed behind the double-slit, see figure~\ref{fig1}.
The mask was moved from one side to the other (figure~\ref{fig2} top to bottom). 
Initially the majority of the electrons are blocked. As the mask is moved, slit 1 becomes partially, then fully open.
When one slit is open, single-slit diffraction can be observed ($\textrm{P}_{1}$ in figure~\ref{fig2} and figure~\ref{fig1}b).
Feynman indicates this as the solid black curve $\textrm{P}_{1}$ (figure~\ref{fig1}b), which is just the central order of the single-slit diffraction pattern.
Because of the finite separation of the mask and double-slit, 
weak double-slit diffraction can be seen in the negative first order of the single-slit diffraction pattern
(see left edge of $\textrm{P}_{1}$ in figure~\ref{fig2}).

As the mask is moved further, more electrons can travel through both slits, changing the pattern from single-slit
to double-slit diffraction. When the mask is centered on the double-slit, both slits are completely open and 
full double-slit diffraction can be observed ($\textrm{P}_{12}$ in figure~\ref{fig2} and figure~\ref{fig1}c). In this position, interaction between the mask 
and the diffracting electrons is negligible. The edges of the mask are 2500 nm away from the center and would
only affect diffraction orders greater than the 60th. The mask is then moved further and the reverse happens; double-slit diffraction
changes back to single-slit diffraction ($\textrm{P}_{2}$ in figure~\ref{fig2} and figure~\ref{fig1}b).
Now, the single-slit diffraction pattern has a weak contribution of double-slit diffraction in its positive first order
(see right edge of $\textrm{P}_{2}$ in figure~\ref{fig2}).
(See Supplementary Movie 1 for more positions of the mask.)

\begin{figure}[tb!]
\centering
\includegraphics[height=.830\textheight]{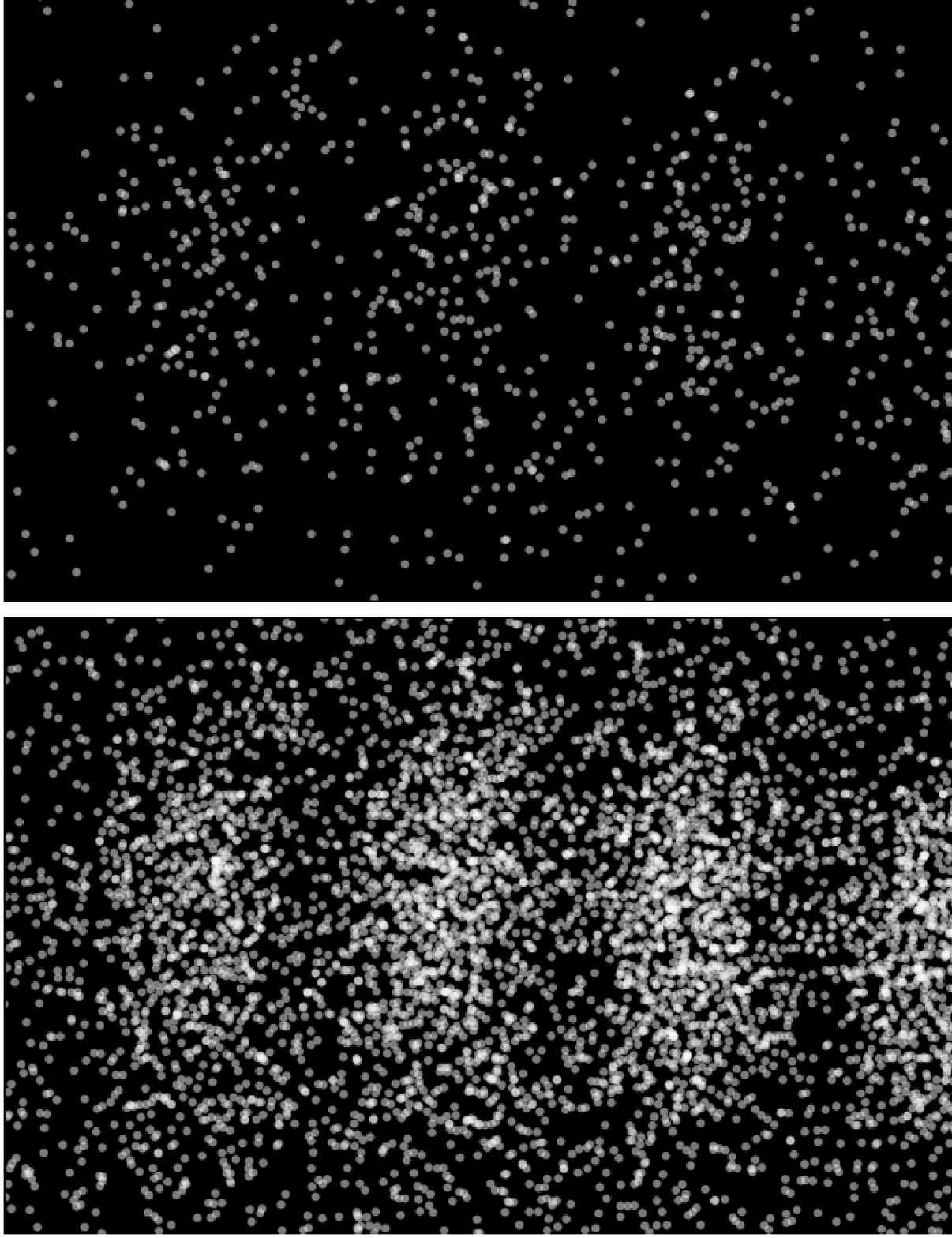}
\caption[Buildup of electron diffraction]{Buildup of electron diffraction. ``Blobs'' indicate the locations of detected electrons.
Shown are intermediate build-up patterns from the central five orders of the diffraction pattern ($\textrm{P}_{12}$) magnified from figure~\ref{fig2},
with 2, 7, 209, 1004, and 6235 electrons (a-e). A full movie of the electron build-up is included in the supplementary data (see Supplementary Movie 2)}
\label{fig3}
\end{figure}

Electron build-up patterns were recorded with the mask centered on the double-slit. The electron source's intensity was reduced
so that the electron detection rate in the pattern was about 1~Hz.  
At this rate and kinetic energy, the average distance between consecutive electrons was $2.3 \, \times \, 10^{6}$~meters.  This ensures
that only one electron is present in the 1 meter long system at any one time, thus eliminating electron-electron interactions.  The
electrostatic quadrupole lens was set to zoom in on the central five diffraction orders. In figure~\ref{fig3} the
build-up of the diffraction pattern is shown. In figure~\ref{fig3}a-c, the electron hits appear to be
completely random and only after many electrons are accumulated can a pattern be discerned, figure~\ref{fig3}d.
In figure~\ref{fig3}e the pattern is clearly visible.
The final build-up of the pattern took about 2 hours.
A full movie of the electron build-up is included in the supplementary data (see Supplementary Movie 2).

\section{Conclusion}
In this paper, we show a full realization of Feynman's thought experiment and illustrate key features of quatum mechanics:
interference and the wave-particle duality of matter.
By controlling the transmission through the individual slits of a double-slit we were able to observe the diffraction patterns from
slit 1 ($\textrm{P}_{1}$), slit 2 ($\textrm{P}_{2}$), and both ($\textrm{P}_{12}$), thus observing the wave properties of electrons.
Also, by recording single electron detection events diffracting through a double-slit we were able to build up a diffraction pattern,
thus observing the particle properties of electrons.

\ack
Roger~Bach and Herman~Batelaan gratefully acknowledge funding from NSF Grant No.~0969506.
Sy-Hwang~Liou acknowledges the support from NSF MRSEC DMR-0820521.
We would like to thank the Nebraska Center for Materials and Nanoscience for the use of their facilities, 
Dr. Jiong Hua for his help in fabricating and taking electron micrographs of the double-slit and mask,
and Xiaolu Yin for help with the coating of the membranes.

\newpage

\newpage

\newpage

\section*{References}
\begin{thebibliography}{99}

\bibitem{Feynman65} Feynman R, Leighton R B and Sands M L 1965 {\it The Feynman Lectures on Physics: Quantum Mechanics} vol~3 (Reading: Addison-Wesley Pub. Co.) ch~1

\bibitem{Jonsson61} J\"{o}nsson C 1961 {Elektroneninterferenzen an mehreren k\"{u}nstlich hergestellten {F}einspalten} {\it Z. Phys} {\bf 161} 454-74

\bibitem{Pozzi76} Merli P G, Missiroli G F and Pozzi G 1976 {On the statistical aspect of electron interference phenomena} {\it Am. J. Phy.} {\bf 44} 306-7

\bibitem{Tonomura89} Tonomura A, Endo J, Matsuda T, Kawasaki T and Ezawa H 1989 {Demonstration of single-electron buildup of an interference pattern} {\it Am. J. Phy.} {\bf 57} 117-20

\bibitem{Beautiful02} Crease R P 2002 {The most beatiful experiment} {\it Phys. World} {\bf 15}(9) 19-20

\bibitem{DoubleSlit02} Crease R P 2002 {The double-slit experiment} {\it Phys. World} {\bf 15}(9) 17 available at http:/\allowbreak /\allowbreak physicsworld.\allowbreak com/\allowbreak cws/\allowbreak article/\allowbreak print/\allowbreak 9745 

\bibitem{Barwick06} Barwick B, Gronniger G, Lu Y, Liou S Y and Batelaan H 2006 {A measurement of electron-wall interaction using transmission diffraction from nanofabricated gratings} {\it J. Appl. Phys.} {\bf 100} 074322

\bibitem{Pozzi07} Frabboni S, Gazzai G C and Pozzi G 2007 {Young's double-slit interference experiment with electrons} {\it Am. J. Phy.} {\bf 75} 1053-5

\bibitem{Pozzi08} Frabboni S, Gazzai G C and Pozzi G 2008 {Nanofabrication and the realization of {Feynman's} two-slit experiment} {\it Appl. Phys. Lett} {\bf 93} 073108

\bibitem{Lindeberg94} Lindeberg T 1994 {\it Scale-Space Theory in Computer Vision} (Dordrecht: Kluwer Academic Publishers)

\bibitem{Lindeberg98} Lindeberg T 1998 {Feature detection with automatic scale selection} {\it Int. J. Comput. Vision} {\bf 30}(2) 79-116

\end{thebibliography}
\end{document}